\documentclass{article}
\PassOptionsToPackage{numbers}{natbib}
\usepackage[preprint]{sty}
\usepackage[utf8]{inputenc} 
\usepackage[T1]{fontenc}    
\usepackage{hyperref}       
\usepackage{url}            
\usepackage{booktabs}       
\usepackage{amsfonts}       
\usepackage{nicefrac}       
\usepackage{microtype}      
\usepackage{xcolor}         
\usepackage{subfig}
\usepackage{amsmath}
\usepackage{amssymb}
\usepackage{algorithm}
\usepackage{algorithmic}
\usepackage{mathtools}
\usepackage{amsthm}
\usepackage{multirow}
\usepackage{pifont}
\usepackage{bm}       
\usepackage{amsmath}  
\usepackage{amssymb}  
\usepackage{amsmath}    
\usepackage{amssymb}    
\usepackage{amsthm}     
\usepackage{hyperref}

\newcommand{\M}{TransMedSeg}

\title{TransMedSeg: A Transferable Semantic Framework for Semi-Supervised Medical Image Segmentation}

\author{
Mengzhu Wang\thanks{Hebei University of Technology}
\And
Jiao Li\thanks{University of Electronic Science and Technology of China}
\And
Shanshan Wang\thanks{Anhui University}
\And
Long Lan\footnotemark[4]
\And
Huibin Tian\footnotemark[4]
\And
Liang Yang\footnotemark[1]
\And
Guoli Yang\thanks{Peking University}
}

\begin{document}

\maketitle

\begin{abstract}
Semi-supervised learning (SSL) has achieved significant progress in medical image segmentation (SSMIS) through effective utilization of limited labeled data. While current SSL methods for medical images predominantly rely on consistency regularization and pseudo-labeling, they often overlook transferable semantic relationships across different clinical domains and imaging modalities. To address this, we propose TransMedSeg, a novel transferable semantic framework for semi-supervised medical image segmentation. Our approach introduces a Transferable Semantic Augmentation (TSA) module, which implicitly enhances feature representations by aligning domain-invariant semantics through cross-domain distribution matching and intra-domain structural preservation. Specifically, TransMedSeg constructs a unified feature space where teacher network features are adaptively augmented towards student network semantics via a lightweight memory module, enabling implicit semantic transformation without explicit data generation. Interestingly, this augmentation is implicitly realized through an expected transferable cross-entropy loss computed over the augmented teacher distribution. An upper bound of the expected loss is theoretically derived and minimized during training, incurring negligible computational overhead. Extensive experiments on medical image datasets demonstrate that TransMedSeg outperforms existing semi-supervised methods, establishing a new direction for transferable representation learning in medical image analysis.

\end{abstract}

\section{Introduction}
Medical image segmentation~\cite{azad2024medical,xing2024segmamba} plays a pivotal role in modern healthcare, enabling precise diagnosis and treatment planning.  While deep learning has revolutionized this field, its success heavily relies on large-scale annotated datasets and poses significant challenges in medical imaging due to the expensive and time-consuming nature of expert annotations. Semi-supervised learning (SSL)~\cite{bortsova2019semi,luo2022semi,wu2022mutual} has emerged as a promising solution to alleviate this annotation bottleneck by leveraging both limited labeled data and abundant unlabeled data. Despite the effectiveness of supervised deep learning methods, their reliance on extensive labeled datasets limits their applicability in many medical imaging scenarios, where expert annotations are scarce and costly to obtain. SSL mitigates this challenge by effectively utilizing the structural and contextual information present in unlabeled data to improve model generalization. Recent advances in semi-supervised medical image segmentation (SSMIS) have explored various strategies, including consistency regularization~\cite{luo2022semi,basak2022exceedingly,wu2022mutual}, pseudo-labeling~\cite{wu2022mutual,wu2023federated,ran2024pseudo}, and adversarial training~\cite{bortsova2019semi,tang2023consistency,xu2025pca}. These approaches encourage the model to learn robust feature representations from unlabeled data while maintaining performance with limited annotations.

\begin{figure*}
	\centering
	\subfloat[GraphCL]{\includegraphics[width=0.3\linewidth]{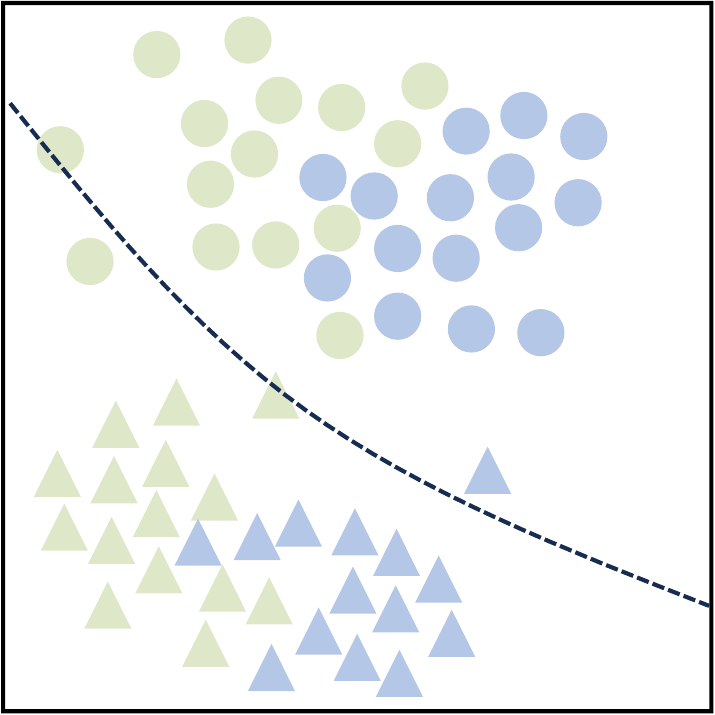}}
        \hspace{1mm}
        \subfloat[\M]{\includegraphics[width=0.3\linewidth]{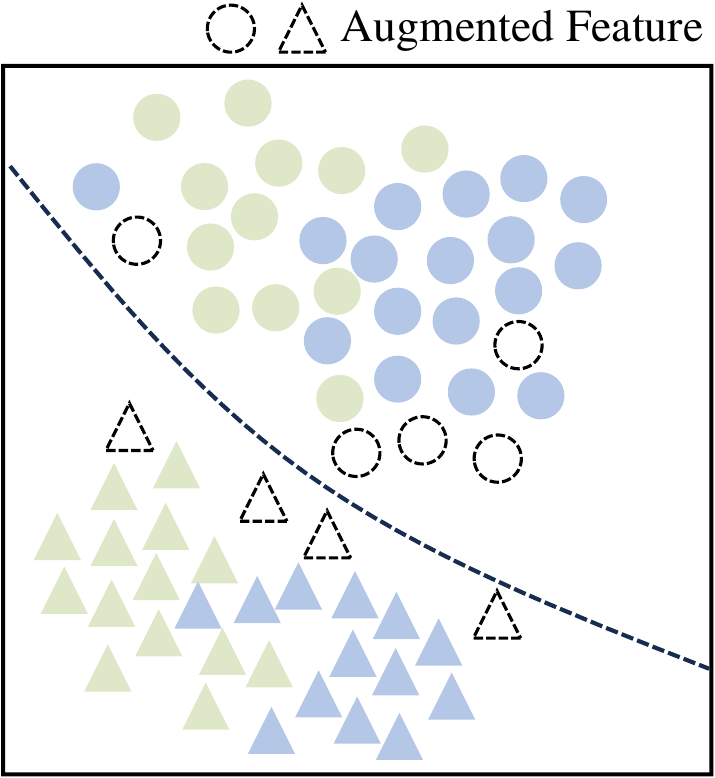}}
	\caption{Overview of TransMedSeg. TransMedSeg enhances the semantic alignment of source features, facilitating the successful adaptation of the final classifier from the student network to the teacher network.}
	\label{figure1}
\end{figure*}

However, current SSL approaches face two fundamental limitations including domain shift in anatomical representations and over-reliance on source-labeled data for feature adaptation that hinder their clinical applicability. These challenges are particularly acute in medical imaging, where variations in scanner protocols, patient populations, and anatomical structures can lead to significant discrepancies between labeled and unlabeled data distributions. The first limitation domain shift in anatomical representations, manifests when models trained on data from one clinical site or scanner type fail to generalize to new environments. For instance, a segmentation model trained on contrast-enhanced CT scans from hospital A may perform poorly on non-contrast scans from hospital B due to differences in image contrast and noise characteristics. The second limitation stems from the heavy dependence on source-labeled data for guiding feature learning. Most SSL approaches generate pseudo-labels or enforce consistency based on predictions from models trained primarily on the limited labeled data. This creates a self-reinforcing cycle where the model's biases from the small labeled set propagate to the unlabeled data. In medical imaging, where class imbalance is extreme (e.g., small lesions against large background regions), this leads to systematic under-segmentation of rare but clinically critical findings.

To address these limitations, we propose a transferable semantic framework for semi-supervised medical image segmentation (TransMedSeg), a novel teacher-student framework that explicitly models cross-domain feature discrepancies while preserving anatomical fidelity. TransMedSeg adopts a teacher-student network architecture, achieving dual optimization of cross-domain feature transfer and anatomical fidelity through an innovative Transferable Semantic Augmentation (TSA) module.  We establish cross-domain feature alignment, where the student network processes labeled source data to extract domain-invariant features while explicitly modeling inter-domain discrepancies through class-specific statistics $(\mu_s^{c}, \Sigma_s^{c})$. Simultaneously, the teacher network progressively adapts to target domain characteristics by aggregating feature statistics $(\mu_t^{c}, \Sigma_t^{c})$ from high-confidence pseudo-labels using exponential moving average updates. The core innovation lies in our Transferable Semantic Augmentation (TSA) module, which performs anatomy-aware feature transformation by sampling from a carefully constructed multivariate normal distribution $\delta \sim \mathcal{N}(\alpha\Delta\mu^{c}, \alpha\Sigma_{t}^{c})$, where $\Delta\mu^{c}$ captures systematic domain shifts and $\Sigma_{t}^{c}$  preserves class-conditional anatomical relationships. Building upon our teacher-student framework with memory-enhanced feature statistics, we further develop an efficient optimization strategy that avoids explicit feature augmentation (as shown in Figure.~\ref{figure1}). Moreover, TransMedSeg enables implicit optimization through an upper-bound loss formulation that avoids the computational overhead of explicit augmentation while providing theoretical guarantees. Further, to validate the effectiveness of our method, we adapt the recent approach GraphCL~\cite{wang2024graphcl} as our baseline. We conducted experiments on multiple datasets, including ACDC~\cite{wu2022exploring}, Pancreas-NIH~\cite{shi2021inconsistency}, and LA~\cite{wu2022exploring}, and the results demonstrate that our method consistently achieves the best performance. The core contributions of this study are summarized as follows:

\begin{itemize}
    \item We develop a feature alignment module that explicitly models domain shifts through teacher-student interaction. By continuously comparing feature distributions between labeled and unlabeled data across institutions, our framework automatically adapts to variations in scanning protocols and patient populations while preserving diagnostic relevance.
    \item The proposed covariance-aware augmentation uniquely incorporates anatomical constraints during feature transformation. This innovation ensures all generated features maintain the physiological plausibility of organ shapes, tissue boundaries, and lesion morphology.
    \item Extensive experiments on popular medical image segmentation benchmarks show that TransMedSeg achieves superior performance.
\end{itemize}

 \begin{figure}[h]
	\centering
	\includegraphics[width=\linewidth]{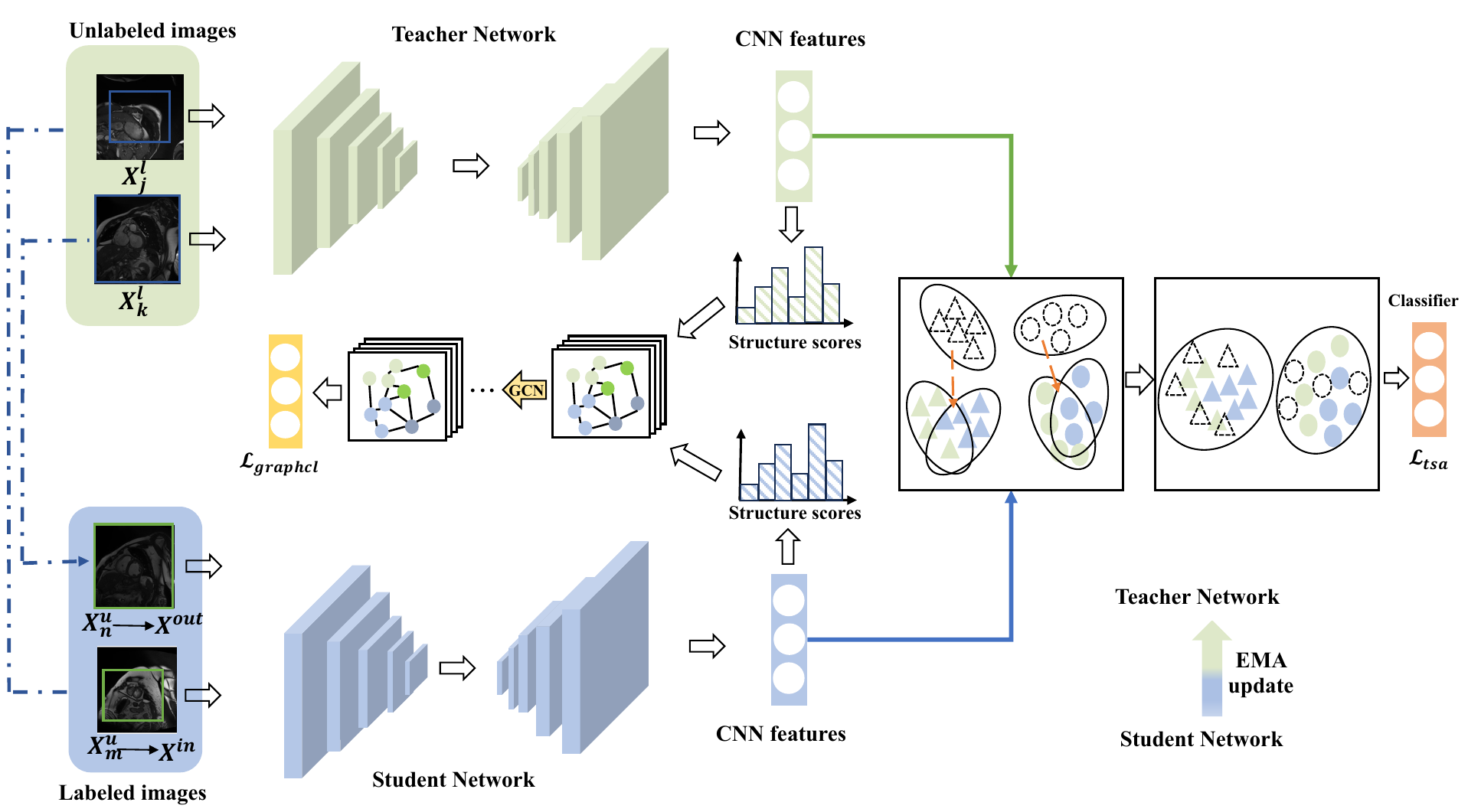}
	
	\caption{Illustration of TransMedSeg. For each class, we leverage the inter-domain feature mean difference (indicated by the orange dashed arrow) and the target intra-class covariance to enhance teacher features, aligning them with the student's style.}
	\label{pipeline}
\end{figure} 

\section{Method}
In semi-supervised medical image segmentation, we process 3D volumetric data represented as $\mathbf{X}\in \mathbb{R}^{C \times W \times H \times D}$, where $C$ indicates imaging modalities and $W \times H \times D$ defines the spatial dimensions. The segmentation task aims to predict voxel-wise labels $\mathbf{Y}\in \{0,1,...,k-1\}^{W \times H \times D}$ for $k$ anatomical structures. Our TransMedSeg framework operates on a dataset $D=\mathcal{D}_{l} \cup \mathcal{D}_{u}$, where $\mathcal{D}_{l} = \{(\mathbf{X}_{i},\mathbf{Y}_{i})\}_{i=1}^{N}$ contains $N$ labeled volumes and $\mathcal{D}_u = \{ \mathbf{X}_j \}_{j=1}^M$ comprises $M \gg N$ unlabeled scans. The proposed teacher-student architecture addresses domain shift through compute class-specific feature discrepancies between teacher and student networks to quantify cross-domain variations and perform anatomy-aware feature augmentation by sampling perturbations $\epsilon \sim \mathcal{N}(\alpha \Delta \mu^c, \alpha \Sigma^c)$, where $\Sigma^c$ captures the student network's estimated intra-class variations.

During training, we create mixed samples by choosing two labeled images \((\mathbf{X}_j^l, \mathbf{X}_k^l)\) and two unlabeled images \((\mathbf{X}_m^u, \mathbf{X}_n^u)\). A foreground region is randomly extracted from \(\mathbf{X}_j^l\) and placed onto \(\mathbf{X}_n^u\) to generate the mixed image \(\mathbf{X}^{\text{out}}\), while another region from \(\mathbf{X}_m^u\) is transferred to \(\mathbf{X}_k^l\) to create \(\mathbf{X}^{\text{in}}\). These blended samples enable the model to acquire rich semantic understanding by incorporating both inward \(\mathbf{X}^{\text{in}}\)and outward \(\mathbf{X}^{\text{out}}\) viewpoints. Our method builds upon a graph-based clustering for semi-supervised medical image segmentation (GraphCL)~\cite{wang2024graphcl}, where GraphCL employ an encoder-decoder architecture. Inspired by transferable semantic augmentation~\cite{wang2021regularizing}. We extend this framework by introducing a novel Transferable Semantic Augmentation (TSA) paradigm that explicitly addresses domain adaptation challenges in medical imaging. While GraphCL effectively captures structural relationships through graph representations, it lacks mechanisms to handle domain shifts commonly encountered in clinical settings. Our TransMedSeg overcomes this limitations through TSA that ensure domain-invariant feature learning and improve generalization from labeled (source) to unlabeled domains. The pipeline of our proposed method is illustrated in Figure~\ref{pipeline}. 

\subsection{Student to Teacher Semantic Augmentation}
Our approach establishes a student to teacher framework for cross-domain adaptation~\cite{wang2024equity}, where the source domain knowledge is distilled through a student network $f_{\text{stu}}(\cdot)$ pre-trained on labeled source data $\mathcal{D}_s = \{(\mathbf{x}_i^s, y_i^s)\}_{i=1}^{N_s}$, while the teacher network $f_{\text{tea}}(\cdot)$ adapts to the unlabeled target domain $\mathcal{D}_t = \{\mathbf{x}_j^t\}_{j=1}^{N_t}$. For each class $c \in \mathcal{C}$, we compute the student's feature space statistics $\bm{\mu}_s^c = \mathbb{E}_{\mathbf{x}^s \sim \mathcal{D}_s^c}[f_{\text{stu}}(\mathbf{x}^s)]$ and $\bm{\Sigma}_s^c = \text{Cov}(f_{\text{stu}}(\mathbf{x}^s))$, where $\mathcal{D}_s^c$ denotes the source samples of class $c$. Correspondingly, the teacher network estimates target domain statistics $\bm{\mu}_t^c$ and $\bm{\Sigma}_t^c$ through online feature aggregation. The core of our method lies in the feature space augmentation strategy that bridges the domain gap. We model the cross-domain transformation as random vectors $\mathbf{z} \sim \mathcal{N}(\Delta\bm{\mu}^c, \bm{\Sigma}_t^c)$, where $\Delta\bm{\mu}^c = \bm{\mu}_s^c - \bm{\mu}_t^c$ captures the inter-domain shift. These vectors generate augmented features  when applied to source samples. The student network is trained to minimize the expected cross-entropy loss over this augmented distribution:

\begin{equation}
\mathcal{L} = \mathbb{E}_{\mathbf{z}}\big[\text{CE}\big(f_{\text{stu}}(\mathbf{x}^s + \mathbf{z}), y^c\big)\big].
\end{equation}

 The teacher network parameters are updated via exponential moving average $\bm{\theta}_{\text{tea}} \leftarrow \lambda\bm{\theta}_{\text{tea}} + (1-\lambda)\bm{\theta}_{\text{stu}}$ to maintain stable feature statistics while gradually adapting to the target domain. This formulation provides three key advantages: (1) the normal distribution sampling efficiently explores meaningful transformation directions without manual design; (2) the class-conditional covariance $\bm{\Sigma}_t^c$ preserves category-aware feature relationships; and (3) the teacher-student interaction ensures robust adaptation while preventing catastrophic forgetting of source knowledge.

\subsection{Transferable Semantic Augmentation (TSA)}
Our framework establishes a dual-network architecture where the student network $f_\text{stu}(\cdot;\theta_s)$, trained on labeled source data $\mathcal{D}_s = \{(\mathbf{x}_i^s,y_i^s)\}_{i=1}^{N_s}$, provides class-discriminative feature representations, while the teacher network $f_\text{tea}(\cdot;\theta_t)$ adapts to unlabeled target data $\mathcal{D}_t = \{\mathbf{x}_j^t\}_{j=1}^{N_t}$. For each class $c \in \{1,...,C\}$, we compute:

\begin{equation}
\bm{\mu}_s^c = \frac{1}{|\mathcal{D}_s^c|}\sum_{\mathbf{x}^s \in \mathcal{D}_s^c} f_\text{stu}(\mathbf{x}^s), \quad
\bm{\Sigma}_s^c = \text{Cov}(f_\text{stu}(\mathbf{x}^s)),
\end{equation}

where $\mathcal{D}_s^c$ denotes source samples of source pixels belonging to class $c$. The teacher network correspondingly estimates target pixel statistics $\bm{\mu}_t^c$ and $\bm{\Sigma}_t^c$ through exponential moving average updates. To address pixel-wise domain shifts in SSMIS, we propose a Transferable Semantic Augmentation (TSA) module that aligns feature at the anatomical structure level. For each class, we compute pixel-wise mean and covariance statistics, where source statistics \((\bm{\mu}^c_s, \bm{\Sigma}^c_s)\) are derived from labeled source-domain pixels and target statistics \((\bm{\mu}^c_t, \bm{\Sigma}^c_t)\) are estimated via exponential moving average (EMA) updates on pseudo-labeled target pixels. The cross-domain transformation is modeled by sampling perturbation vectors from a class-conditional normal distribution:

\begin{equation}
\bm{z}^c \sim \mathcal{N}(\underbrace{\bm{\mu}_t^c - \bm{\mu}_s^c}_{\Delta\bm{\mu}^c}, \bm{\Sigma}_t^c),
\end{equation}
where the inter-domain mean different $\Delta\bm{\mu}^c$ captures systematic shifts, while $\bm{\Sigma}_t^c$ models intra-class variations.

For each source-domain pixel embedding $\mathbf{f}_{s,i}^c \in \mathbb{R}^d$ belonging to class $c=y_{si}$, we generate augmented features through multivariate Gaussian sampling $\mathcal{T}^c \sim \mathcal{N}(\alpha\boldsymbol{\Delta\mu}^c, \alpha\boldsymbol{\Sigma}_t^c)$, which $\alpha \in [0,1]$ controls adaptation strength. The augmented pixel feature is generated via $\tilde{\mathbf{f}}_{s,i}^c = \mathbf{f}_{s,i}^c + \delta$ and $\delta \sim \mathcal{T}^c$. To get more navie explicit augmentation, we reformulate the feature augmentation process to maintain anatomical consistency while achieving computational efficiency. For each source pixel feature $\mathbf{f}_{s,t}^c \in \mathbb{R}^d \; (\textrm{class } c = y_{st})$, generated $M$ times with its labeled preserved, which will result in an augmented feature:
\begin{equation}
\mathcal{F}_{\text{aug}} = \bigcup_{i=1}^{n_s}\bigcup_{m=1}^M \left\{\left({f}_{s,i}^c + \delta_m, y_{si}\right) \mid \delta_m \sim \mathcal{N}(\alpha\boldsymbol{\Delta\mu}^c, \alpha\boldsymbol{\Sigma}_t^c)\right\},
\end{equation}
with the empirical segmentation loss:
\begin{align}
\mathcal{L}_\text{tsa} &= \frac{1}{n_sM}\sum_{i=1}^{n_s}\sum_{m=1}^M \ell(\mathbf{f}_{s,i}^c + \delta_m, y_{si}), \\
\ell(\mathbf{f},y) &= -\log\left(\frac{\exp(\boldsymbol{w}y^\top\mathbf{f} + b_y)}{\sum_{c=1}^C \exp(\boldsymbol{w}_c^\top\mathbf{f} + b_c)}\right),
\end{align}
where $n_s$ denotes the total number of pixels in the source domain image. $\boldsymbol{w}_c \in \mathbb{R}^d$ and $\boldsymbol{b}_c \in \mathbb{R}^d$ denote the classifier weight vector and bias term for class $c$, respectively.

Traditional explicit augmentation methods require sampling multiple perturbed versions of each pixel feature to achieve stable convergence, leading to a high memory cost. Different from these strategies, our method circumvents this through implicit semantic data augmentation (ISDA)~\cite{wang2021regularizing}. By leveraging ISDA, our approach avoids the need for explicit sampling while maintaining the benefits of robust feature regularization. Specifically, we model the augmentation process as a continuous transformation space, where each pixel feature is enriched through an implicit distribution. This allows the network to effectively approximate an infinite set of augmented views. Then when $M \to \infty$, we establish an upper-bound loss through asymptotic analysis of the augmented source distribution. By the Strong Law of Large Numbers~\cite{etemadi1981elementary}, the empirical risk converges almost surely to its expectation.
\begin{equation}
\label{Eq:transferable_CE_loss}
\begin{aligned}
 \lim_{M\to\infty} \mathcal{L}_M 
&= \frac{1}{n_s} \sum_{i=1}^{n_s} \mathbb{E}_{\tilde{\mathbf{f}}_{si}} \left[ -\log \left( \frac{\exp(\boldsymbol{w}_{y_{si}}^\top \tilde{\mathbf{f}}_{si} + b_{y_{si}})} {\sum_{c=1}^C \exp(\boldsymbol{w}_c^\top \tilde{\mathbf{f}}_{si} + b_c)} \right) \right] \\
& \leq \frac{1}{n_s}\sum_{i=1}^{n_s}\log\mathbb{E}\left[\sum_{c=1}^C \exp\left(\Delta\boldsymbol{w}_c^\top\tilde{\mathbf{f}}_{si} + \Delta b_c\right)\right].
\end{aligned}
\end{equation}

Direct optimization of Eq.~\eqref{Eq:transferable_CE_loss} is computationally prohibitive due to the expectation over infinite augmentations. By applying Jensen's inequality \cite{mcshane1937jensen} to the concave log function, we obtain $\mathbb{E}[\log(X)] \leq \log(\mathbb{E}[X])$. This leads to the following upper bound derivation:
\begin{equation}
\begin{aligned}
\label{js}
\mathcal{L}_\text{tsa} &\triangleq \lim_{M\to\infty} \mathcal{L}_M \\
&\leq \frac{1}{n_s}\sum_{i=1}^{n_s}\log\mathbb{E}\left[\sum_{c=1}^C \exp\left(\Delta\boldsymbol{w}_c^\top\tilde{\mathbf{f}}_{si} + \Delta b_c\right)\right] \\
&= \frac{1}{n_s}\sum_{i=1}^{n_s}\log\sum_{c=1}^C \mathbb{E}\left[\exp\left(\Delta\boldsymbol{w}_c^\top\tilde{\mathbf{f}}_{si} + \Delta b_c\right)\right],
\end{aligned}
\end{equation}
where $\Delta\boldsymbol{w}_c \equiv \boldsymbol{w}_{y_{si}}$ and $\Delta b_c \equiv b_c - b_{y_s}$. The surrogate loss $\mathcal{L}_{\infty}$ in Eq.~\eqref{js} provides a theoretically grounded framework for SSMIS. Moreover, the proposed novel loss allows us to utilize TSA as a plug-in module for other SSMIS methods to further improve their transferability.

\subsection{Overall Formulation}
Our method builds upon the GraphCL framework~\cite{wang2024graphcl}, which leverages graph-based clustering and semi-supervised learning to improve medical image segmentation. The approach is designed to effectively utilize both labeled and unlabeled data by integrating structural information through graph neural networks. This is achieved by combining foreground regions from labeled images with background regions from unlabeled images, and vice versa, using a binary mask. These mixed samples are then used to train a teacher-student model, where the teacher generates pseudo-labels for unlabeled data, and the student learns from both the pseudo-labels and the ground-truth annotations. The loss function in this framework combines Dice and cross-entropy losses to optimize segmentation performance. 

Further, GraphCL constructs a graph where each node represents an image region, and edges are weighted based on feature similarity. The adjacency matrix of this graph is derived from both CNN-extracted features and structure-aware scores. A Graph Convolutional Network (GCN) is then applied to propagate features across the graph, enabling the model to learn from both local and global structural patterns. This graph-based approach ensures that the segmentation model can better exploit contextual relationships, particularly in medical images where anatomical structures exhibit complex spatial dependencies. These methods lack transferability across different clinical domains and imaging modalities. Inspired by the transferable semantic augmentation paradigm from domain adaptation research, we propose transferable semantic augmentation. Specifically, our final loss function is formulated as follows:
\begin{equation}
\begin{aligned}
\label{overal formulation}
\mathcal{L}_\text{TransMedSeg} = \mathcal{L}_\text{GraphCL} + \beta \mathcal{L}_\text{tsa}.
\end{aligned}
\end{equation}
We use a weight $\beta$ to control transferable semantic augmentation in the loss function. The training process of our TransMedSeg is summarized in Appendix B.

\section{Experiments}

\subsection{Experimental Settings}
\noindent\textbf{Datasets.} 
To validate the effectiveness of \M, we conduct extensive experiments on three benchmark datasets. 
\textbf{ACDC} dataset is a cardiac MRI segmentation dataset comprising short-axis cine-MR images from 100 patients, annotated with four semantic classes: background, right ventricle, left ventricle, and myocardium. The dataset is split in a fixed manner, with 70 cases used for training, 10 for validation, and 20 for testing~\cite{wu2022exploring}.
\textbf{Pancreas-NIH} dataset is a publicly available dataset consisting of 82 contrast-enhanced abdominal CT scans with expert manual annotations of the pancreas. Each 3D volume has a fixed in-plane resolution of \(512 \times 512\), with the number of slices \(D \in [181, 466]\). For preprocessing, we apply the soft-tissue window \([-120, 240]\) Hounsfield units, crop the volumes around the pancreas region with 25-voxel margins, and randomly crop \(96 \times 96 \times 96\) patches during training. In conclusion, we follow the setting in CoraNet~\cite{shi2021inconsistency} and BCP~\cite{bai2023bidirectional} for fair comparison.
\textbf{LA} dataset is a cardiac MRI segmentation dataset released by the 2018 Atrial Segmentation Challenge, consisting of 100 3D gadolinium-enhanced MR imaging volumes with expert annotations for the left atrial cavity. Each scan has an isotropic resolution of \(0.625 \times 0.625 \times 0.625\,\mathrm{mm}^3\), though the spatial dimensions vary across subjects. Following prior works~\cite{shi2021inconsistency, bai2023bidirectional}, we adopt the fixed data split with 80 scans for training and 20 for testing. For semi-supervised settings, 4 scans (5\%) are selected as labeled data, and the remaining training scans are treated as unlabeled.

\noindent\textbf{Metrics.} Following prior studies~\cite{luo2021semi, zhao2024alternate}, we evaluate the segmentation performance using four widely adopted metrics: Dice Similarity Coefficient (Dice), Jaccard Similarity Coefficient (Jaccard), 95\% Hausdorff Distance (95HD), and Average Surface Distance (ASD). Dice and Jaccard are region-based metrics that assess the overlap between the predicted segmentation and ground truth, with higher values indicating better performance. In contrast, 95HD and ASD are boundary-based metrics that measure the contour discrepancy between predictions and annotations, where lower values represent better agreement.

\subsection{Experimental Results}
\noindent\textbf{Performance Comparison.} 
We evaluated \M~against state-of-the-art semi-supervised methods on three benchmark datasets. These experiments test the model's ability to handle domain shifts and leverage unlabeled data effectively across different modalities and anatomical structures.
On the \textbf{LA} dataset (Table~\ref{tab:la}), using only 8 (10\%) labeled data, \M~achieves the best performance in Dice (89.62\%), Jaccard (81.31\%), and 95HD (6.68 voxel), slightly outperforming the strong GraphCL baseline and significantly improving upon other methods like UA-MT and the supervised V-Net. Even with fewer labels (5\%), \M~leads in Dice (87.65\%) and Jaccard (78.24\%), likely aided by the Transferable Semantic Augmentation (TSA) module.
On the \textbf{ACDC} dataset (Table~\ref{tab:acdc}), \M~again shows superior performance. With 7 (10\%) labeled data, it achieves top scores across all metrics: Dice (89.96\%), Jaccard (82.31\%), 95HD (1.61 voxel), and ASD (0.64 voxel). This represents a notable improvement over GraphCL. The particularly strong performance in boundary metrics (95HD, ASD) suggests that the TSA effectively preserves the complex and variable shapes of cardiac structures while adapting to the target domain, mitigating the over-reliance on potentially biased source labels.  With only 3 (5\%) labeled data further emphasizing its ability to leverage unlabeled data efficiently.
On the \textbf{Pancreas-NIH} dataset (Table~\ref{tab:pan}), which uses 12 (20\%) labeled data, \M~continues its leading performance, achieving the best Dice (83.06\%), Jaccard (71.26\%), 95HD (5.67 voxel), and ASD (1.51 voxel). It surpasses all baselines. The pancreas segmentation task is more challenging due to high anatomical variability and often subtle boundaries in CT data. The superior performance, especially in boundary metrics (95HD, ASD), underscores the effectiveness of \M~'s covariance-aware augmentation in maintaining anatomical fidelity specific to the target CT data.
Across all three datasets, representing different imaging modalities (MRI, CT), anatomical regions, and labeled data ratios, \M~consistently delivers state-of-the-art results. The specific improvements observed on each dataset validate the important contributions of \M.

\begin{table}
    \centering
    
    \caption{Quantitative comparisons with state-of-the-art semi-supervised segmentation methods on \textbf{LA} dataset. The best results are in bold.} 
    \setlength{\tabcolsep}{1pt} 
    \renewcommand{\arraystretch}{1} 
    \resizebox{\columnwidth}{!}{ 
    \begin{tabular}{c|cc|cccc|cc|cccc}
        \hline  
        \multirow{2}{*}{Method} & \multicolumn{2}{c|}{Scans used} & \multicolumn{4}{c|}{Metrics} & \multicolumn{2}{c|}{Scans used} & \multicolumn{4}{c}{Metrics}\\  
        \cline{2-3} \cline{4-7}  \cline{8-9} \cline{10-13}
        & Labeled & Unlabeled & Dice$\uparrow$ & Jaccard$\uparrow$ & 95HD$\downarrow$ & ASD$\downarrow$ & Labeled & Unlabeled & Dice$\uparrow$ & Jaccard$\uparrow$ & 95HD$\downarrow$ & ASD$\downarrow$\\  
        \hline  
        V-Net~\cite{milletari2016v} & 4(5\%) & 0 & 52.55 & 39.69 & 47.05 & 9.87 & 8(10\%) & 0 & 82.74 & 71.72 & 13.35 & 3.26 \\  
        \hline  
        UA-MT~\cite{yu2019uncertainty} & & & 82.26 & 70.98 & 13.71 & 3.82 & & & 87.79 & 78.39 & 8.68 & 2.12 \\  
        SASSNet~\cite{li2020shape} & & & 81.60 & 69.63 & 16.16 & 3.58 & & & 87.54 & 78.05 & 9.84 & 2.59 \\  
        DTC~\cite{luo2021semi} & & & 81.25 & 69.33 & 14.90 & 3.99 & & & 87.51 & 78.17 & 8.23 & 2.36 \\  
        URPC~\cite{luo2021efficient} & \multirow{2}{*}{4(5\%)}  & \multirow{2}{*}{76(95\%)}  & 82.48 & 71.35 & 14.65 & 3.65& \multirow{2}{*}{8(10\%)}  & \multirow{2}{*}{72(90\%)} & 86.92 & 77.03 & 11.13 & 2.28 \\
        MC-Net~\cite{wu2021semi} & & & 83.59 & 72.36 & 14.07 & 2.70 & & & 87.62 & 78.25 & 10.03 & 1.82 \\  
        SS-Net~\cite{wu2022exploring} & & & 86.33 & 76.15 & 9.97 & 2.31 & & & 88.55 & 79.62 & 7.49 & 1.90 \\  
        BCP~\cite{bai2023bidirectional} & & & 87.06 & 77.42 & 8.83 & \textbf{2.15} & & & 89.39 & 80.92 & 7.26 & 1.76 \\   
        GraphCL~\cite{wang2024graphcl} & & & 86.55 & 76.52 & 10.64 & 2.70 & & & 89.49 & 81.09 & 7.45 & 1.76 \\  
        \hline
        \M & & & \textbf{87.65} & \textbf{78.24} & \textbf{8.71} & 2.25 & & & \textbf{89.62} & \textbf{81.31} & \textbf{6.68} & \textbf{1.75} \\
        \hline
    \end{tabular}}
    \label{tab:la}
\end{table}

\begin{table}
    \centering
    \caption{Quantitative comparisons with state-of-the-art semi-supervised segmentation methods on \textbf{ACDC} dataset. The best results are in bold.}
    \setlength{\tabcolsep}{1pt} 
    \renewcommand{\arraystretch}{1} 
    \resizebox{\columnwidth}{!}{ 
    \begin{tabular}{c|cc|cccc|cc|cccc}
        \hline  
        \multirow{2}{*}{Method} & \multicolumn{2}{c|}{Scans used} & \multicolumn{4}{c|}{Metrics} & \multicolumn{2}{c|}{Scans used} & \multicolumn{4}{c}{Metrics}\\  
        \cline{2-3} \cline{4-7}  \cline{8-9} \cline{10-13}
        & Labeled & Unlabeled & Dice$\uparrow$ & Jaccard$\uparrow$ & 95HD$\downarrow$ & ASD$\downarrow$ & Labeled & Unlabeled & Dice$\uparrow$ & Jaccard$\uparrow$ & 95HD$\downarrow$ & ASD$\downarrow$\\  
        \hline  
        U-Net~\cite{ronneberger2015u} & 3(5\%) &0 & 47.83 & 37.01 & 31.16 & 12.62 & 7(10\%) &0 & 79.41 & 68.11 & 9.35 & 2.70 \\

        \hline  
        UA-MT~\cite{yu2019uncertainty} & & & 46.04 & 35.97 & 20.08 & 7.75 & & & 81.65 & 70.64 & 6.88 & 2.02 \\
        SASSNet~\cite{li2020shape} & & & 57.77 & 46.14 & 20.05 & 6.06 & & & 84.50 & 74.34 & 5.42 & 1.86 \\
        DTC~\cite{luo2021semi} & & & 56.90 & 45.67 & 23.36 & 7.39 & & & 84.29 & 73.92 & 12.81 & 4.01 \\
        URPC~\cite{luo2021efficient} & \multirow{2}{*}{3(5\%)} & \multirow{2}{*}{67(95\%)} & 55.87 & 44.64 & 13.60 & 3.74 & \multirow{2}{*}{7(10\%)} & \multirow{2}{*}{63(90\%)} & 83.10 & 72.41 & 4.84 & 1.53 \\
        MC-Net~\cite{wu2021semi} & & & 62.85 & 52.29 & 7.62 & 2.33 & & & 86.44 & 77.04 & 5.50 & 1.84 \\
        SS-Net~\cite{wu2022exploring}  & & & 65.83 & 55.38 & 6.67 & 2.28 & & & 86.78 & 77.67 & 6.07 & 1.40 \\
        BCP~\cite{bai2023bidirectional}  & & & 86.83 &77.64 & 8.71 & 2.47 & & & 88.84 &80.61 & 4.42& 1.38 \\
        
        GraphCL~\cite{wang2024graphcl} & & & 87.96 & 79.26 & 4.10 & 1.28 & & & 88.92 & 80.85 & 3.90 & 1.04 \\ 
        \hline
        \M & & & \textbf{88.14} & \textbf{79.48} & \textbf{3.42} & \textbf{0.97} & & & \textbf{89.96} & \textbf{82.31} & \textbf{1.61} & \textbf{0.64} \\  
        \hline
    \end{tabular}
    }
    \label{tab:acdc}
\end{table}  

\begin{table}
    \centering
    \caption{Quantitative comparisons with state-of-the-art semi-supervised segmentation methods on \textbf{Pancreas-NIH} dataset. The best results are in bold.}
    
    \setlength{\tabcolsep}{3pt} 
    \renewcommand{\arraystretch}{1} 
    \resizebox{0.65\columnwidth}{!}{ 
    \begin{tabular}{c|cc|cccc}  
        \hline  
        \multirow{2}{*}{Method} & \multicolumn{2}{c|}{Scans used} & \multicolumn{4}{c}{Metrics} \\  
        \cline{2-3} \cline{4-7}  
        & Labeled & Unlabeled & Dice$\uparrow$ & Jaccard$\uparrow$ & 95HD$\downarrow$ & ASD$\downarrow$ \\ 
        \hline 
        V-Net~\cite{milletari2016v} & \multirow{9}{*}{12(20\%)}  & \multirow{9}{*}{50(80\%)} & 69.96 & 55.55 & 14.27 & 1.64 \\
        DAN~\cite{zhang2017deep} & & & 76.74 & 63.29 & 11.13 & 2.97 \\
        ADVNET~\cite{vu2019advent} & & & 75.31 & 61.73 & 11.72 & 3.88 \\
        UA-MT~\cite{yu2019uncertainty} & & & 77.26 & 63.82 & 11.90 & 3.06 \\
        SASSNet~\cite{li2020shape} & & & 77.66 & 64.08 & 10.93 & 3.05 \\
        DTC~\cite{luo2021semi}  & & & 78.27 & 64.75 & 8.36 & 2.25 \\
        CoraNet~\cite{shi2021inconsistency} & & & 79.67 & 66.69 & 7.59 & 1.89 \\
        BCP~\cite{bai2023bidirectional}  & & & 81.12 & 68.81 &8.11 & 2.34 \\ 
        
        GraphCL~\cite{wang2024graphcl} & & & 82.78 &70.84 & 8.00 & 2.86  \\
        \hline
        \M & & & \textbf{83.06} &\textbf{71.26} & \textbf{5.67}  & \textbf{1.51}  \\
        \hline  
    \end{tabular}}
    \label{tab:pan}
\end{table}

\begin{figure*}[htpb]
    \centering
    \subfloat[w/o $\mathcal{L}_{\text{tsa}}$]{\includegraphics[width=0.23\columnwidth]{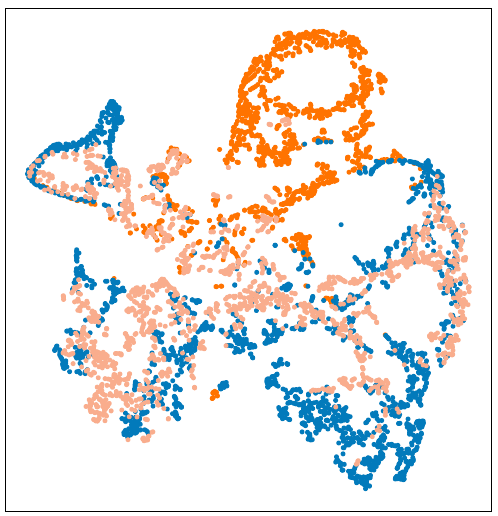}}
    \subfloat[\M]{\includegraphics[width=0.23\columnwidth]{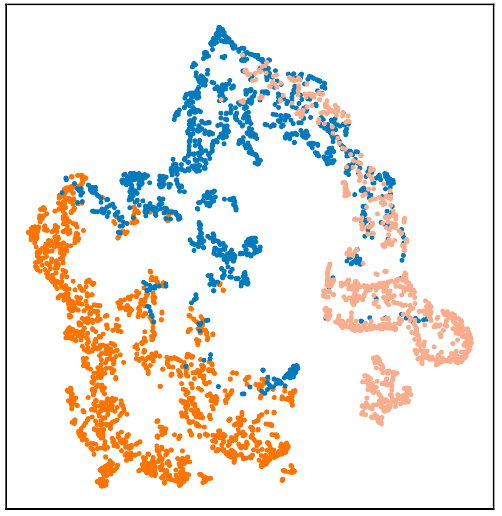}}
    \subfloat[$\beta$ on ACDC(5\%)]{\includegraphics[width=0.28\columnwidth]{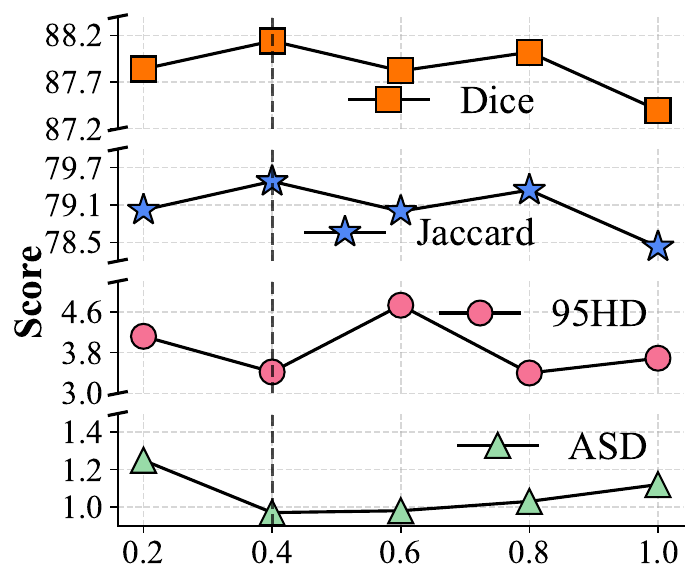}}
    \subfloat[$\beta$ on ACDC(10\%)]{\includegraphics[width=0.27\columnwidth]{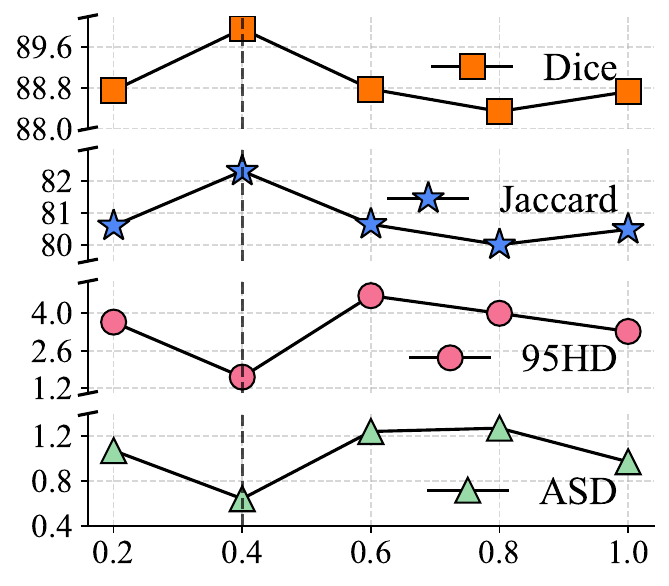}}
    \caption{(a–b) T-SNE visualization of feature embeddings w/o and with $\mathcal{L}_{\text{tsa}}$. (c-d) Sensitivity analysis of $\mathcal{L}_{\text{tsa}}$ with respect to the weighting coefficient $\beta$ on ACDC dataset.}
    \label{fig:sensitivity_tsne}
\end{figure*}

\noindent\textbf{Ablation Study.} 
To quantitatively evaluate the effect of the proposed $\mathcal{L}_{\text{tsa}}$, we conduct ablation experiments on the ACDC and Pancreas-NIH datasets under different labeled/unlabeled splits. As reported in Table~\ref{tab:ablation_two}, removing $\mathcal{L}_{\text{tsa}}$ consistently leads to performance drops across all metrics, highlighting its essential role in improving segmentation quality under limited supervision. Specifically, on ACDC with only 5\% labeled data, adding $\mathcal{L}_{\text{tsa}}$ improves Dice from 87.96\% to 88.14\%, Jaccard from 79.26\% to 79.48\% and reduces 95HD from 4.10 to 3.42, ASD from 1.28 to 0.97. When the supervision increases to 10\%, the gains become more substantial. Similar trends are observed on Pancreas-NIH. These results validate the effectiveness of our dual-phase semantic adaptation strategy.
To qualitatively evaluate the effect of \(\mathcal{L}_{\text{tsa}}\), we also visualize the feature embeddings using t-SNE in Figure~\ref{fig:sensitivity_tsne}(a–b). With \(\mathcal{L}_{\text{tsa}}\) enabled (Figure~\ref{fig:sensitivity_tsne}(b)), the features form well-separated and compact clusters with clear inter-domain alignment. In contrast, removing \(\mathcal{L}_{\text{tsa}}\) (Figure~\ref{fig:sensitivity_tsne}(a)) results in significant overlap and entanglement across domains. These results confirm that \(\mathcal{L}_{\text{tsa}}\) reduces domain discrepancy and enhances semantic consistency across domains.

\begin{table*}[t]
    \centering
    \caption{Ablation experiments on ACDC and Pancreas-NIH datasets.}
    \small
    \setlength{\tabcolsep}{2pt} 
    \renewcommand{\arraystretch}{0.9} 
    \begin{tabular}{c|cc|cc|cccc}  
        \hline
        \multirow{2}{*}{Dataset} & \multirow{2}{*}{Baseline} & \multirow{2}{*}{\(\mathcal{L}_{tsa}\)} & \multicolumn{2}{c|}{Scans used} & \multicolumn{4}{c}{Metrics} \\
        \cline{4-5} \cline{6-9}
        & & & Labeled & Unlabeled & Dice$\uparrow$ & Jaccard$\uparrow$ & 95HD$\downarrow$ & ASD$\downarrow$ \\ 
        \hline 
        \multirow{4}{*}{ACDC} & \ding{52} & \ding{56} & \multirow{2}{*}{5\%} & \multirow{2}{*}{95\%} & 87.96 & 79.26 & 4.10 & 1.28 \\
        & \ding{52} & \ding{52} & & & \textbf{88.14} & \textbf{79.48} & \textbf{3.42} & \textbf{0.97} \\
        \cline{2-9} 
         & \ding{52} & \ding{56} & \multirow{2}{*}{10\%} & \multirow{2}{*}{90\%} & 88.92 & 80.85 & 3.90 & 1.04 \\
        & \ding{52} & \ding{52} & & & \textbf{89.96} & \textbf{82.31} & \textbf{1.61} & \textbf{0.64} \\
        \hline
        \multirow{2}{*}{Pancreas} & \ding{52} & \ding{56} & \multirow{2}{*}{20\%} & \multirow{2}{*}{80\%} & 82.78 &70.84 & 8.00 & 2.86 \\
        & \ding{52} & \ding{52} & & & \textbf{83.06} &\textbf{71.26} & \textbf{5.67}  & \textbf{1.51} \\
        \hline
    \end{tabular}
    \label{tab:ablation_two}
\end{table*}

\noindent\textbf{Sensitivity Analysis.} 
We perform a sensitivity analysis on the weighting coefficient $\beta$ of \(\mathcal{L}_{\text{tsa}}\) using the ACDC dataset with 5\% and 10\% labeled data, as illustrated in Figure~\ref{fig:sensitivity_tsne}(c-d). For $\beta$, the results demonstrate that the segmentation metrics remain relatively stable over the range of 0.2 to 1.0, and all metrics perform best at 0.4.
Therefore, $\beta$ is set to 0.4 in our experiments according to the exact result.

\begin{figure*}
	\centering
	\subfloat[Visualizations on ACDC(5\%)]{\includegraphics[width=0.45\linewidth]{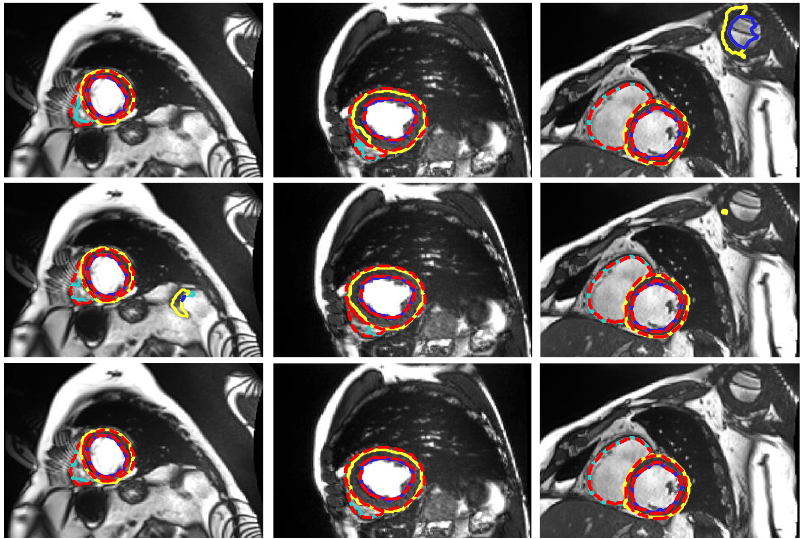}}
        \subfloat[Visualizations on ACDC(10\%)]{\includegraphics[width=0.55\linewidth]{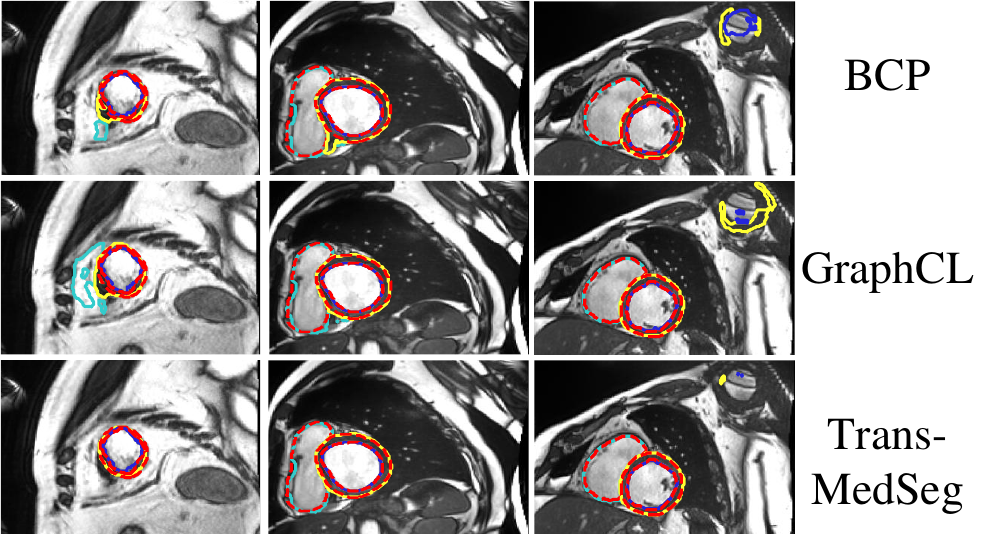}}
	\caption{Qualitative visualization results from the ACDC dataset.}
	\label{fig:acdc_visual}
\end{figure*}

\begin{figure*}
	\centering
        \includegraphics[width=1.0\linewidth]{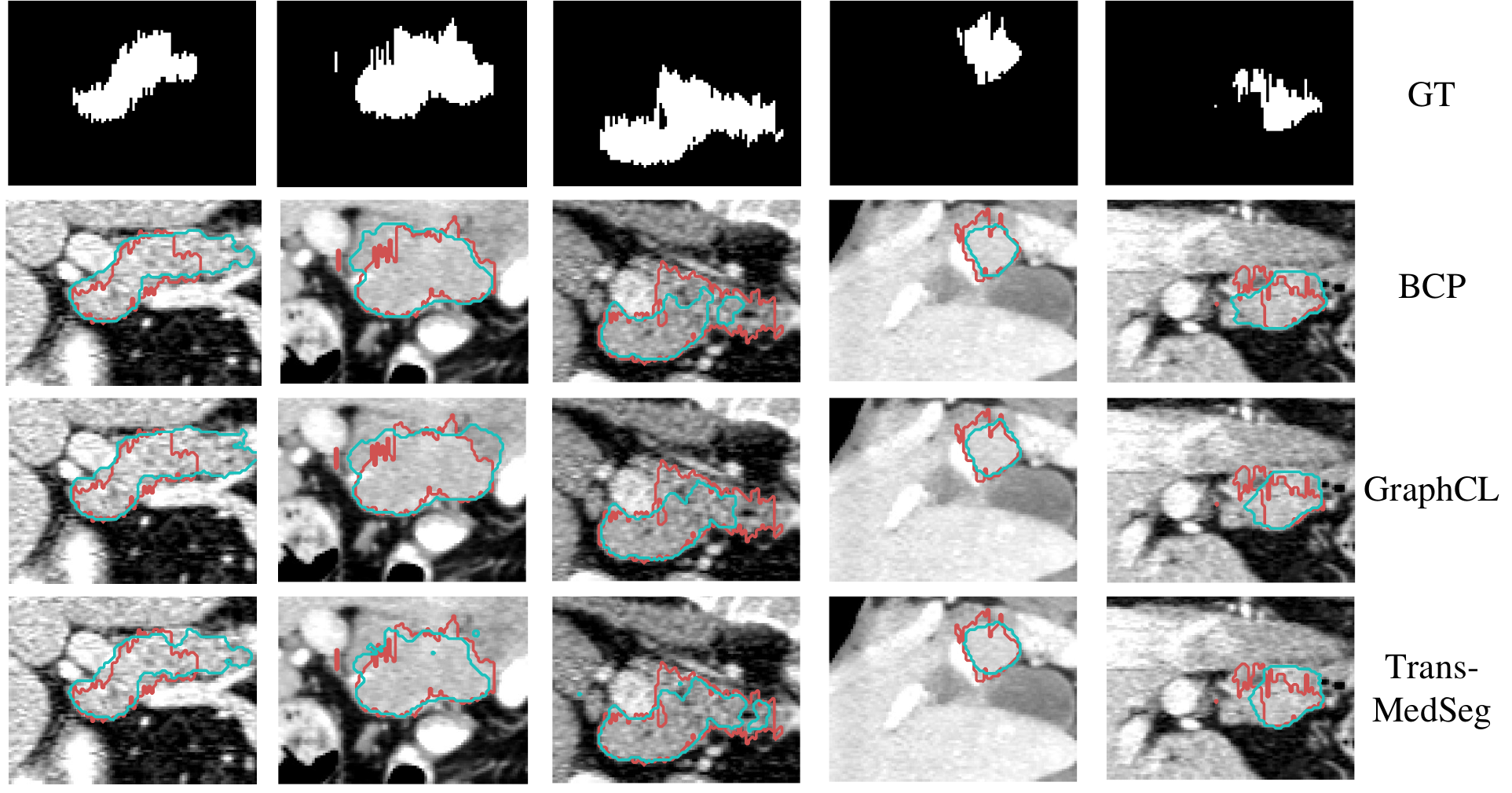}
	\caption{Qualitative visualization results from the Pancreas-NIH dataset.}
	\label{fig:pan_visual}
\end{figure*}

\noindent\textbf{Qualitative Visualization.} 
Figure~\ref{fig:acdc_visual} presents qualitative segmentation results on the ACDC dataset under semi-supervised settings with 5\% and 10\% labeled data. Each row corresponds to a different method: BCP (top row), GraphCL (middle row), and our proposed \M~(bottom row). The red contours denote the ground truth labels, while the other colored contours show the predicted segmentations. It is evident that our \M~method produces segmentation results that more closely align with the ground truth compared to BCP and GraphCL across various cardiac structures. For instance, in both 5\% and 10\% labeled cases, \M~ correctly delineates the cardiac chambers and boundaries with fewer errors and mis-segmentations. In contrast, BCP and GraphCL exhibit more inconsistencies and uncertain boundaries, especially at domain transition regions.
Figure~\ref{fig:pan_visual} presents qualitative segmentation results on the Pancreas-NIH dataset, comparing ground truth (GT) annotations with predictions from BCP, GraphCL, and our proposed \M~. Red contours represent GT boundaries, while green contours indicate the predicted segmentations. In the first and second columns, \M~ demonstrates improved delineation of fine structural details and better alignment with GT contours. Specifically, in the first case, BCP and GraphCL slightly under-segment the pancreatic head, whereas \M~ effectively mitigates this issue. The third column illustrates a challenging case with irregular morphology and low contrast. In this scenario, both BCP and GraphCL exhibit noticeable segmentation errors, particularly at the borders, while \M~ produces a more precise and consistent contour.

\section{Conclusion}

In this paper, we presented TransMedSeg, a novel transferable semantic framework for semi-supervised medical image segmentation (SSMIS). The proposed framework addresses two critical limitations of existing SSL methods: domain shift in anatomical representations and over-reliance on source-labeled data. By introducing a Transferable Semantic Augmentation (TSA) mechanism, TransMedSeg aligns domain-invariant semantics through cross-domain distribution matching and intra-domain structural preservation, enabling robust feature learning across diverse clinical domains and imaging modalities. The teacher-student architecture, coupled with a lightweight memory module, facilitates implicit semantic transformation without explicit data generation, optimizing both computational efficiency and performance. Extensive experiments on benchmark datasets (ACDC, Pancreas-NIH, and LA) demonstrate that TransMedSeg consistently outperforms state-of-the-art semi-supervised medical image segmentation. In the future, we will integrate advanced techniques with this approach to further minimize annotation costs, particularly for dense-annotation-dependent multi-organ segmentation tasks.

\appendix

\clearpage
\bibliographystyle{plainnat}
\bibliography{reference}

\end{document}